# Integrated scalable metalens array for optical addressing

Tie Hu,[1] YunXuan Wei,[1] Zhihao Zhou,[1] Ming Zhao,[1*] and ZhenYu Yang[1*]

[1] *School of Optical and Electronic Information, Huazhong University of Science and Technology, Wuhan 430074, China*

*Corresponding author: zhaoming@hust.edu.cn*

**Abstract:** The scalability of optical addressing is one of the key challenges for the large-scale trapped-ion quantum computing (QC). Metasurfaces is a promising solution due to excellent optical performance, integration, and miniaturization. Here, we propose and numerically demonstrate a design of integrated scalable metalens array (ISMA) working in the UV 350nm range for optical addressing. These ISMAs can focus the collimated individual addressing beams array into diffraction-limited spots array, featuring crosstalk below 0.82%, focusing efficiency as high as 36.90%. For various applications, the x-polarized, polarization-insensitive and right-handed circularly polarized ISMA are designed. The design proposed in this paper can also find applications in optical trapping.

1. Introduction

The technology to focus and align the individual addressing beams to the trapped ion is called optical addressing. Currently, the scalability of optical addressing is mainly subject to its focusing structure, thus the focusing structure with high performance as well as good scalability is vital for large-scale trapped-ion QC. Previously, there are many pioneering works about the focusing structures for optical addressing, like the refractive elements based on free-space optics[1, 2], the microfabricated mirrors[3, 4], fibers attached to an ion trap[5, 6] and the focusing grating couplers[7-9], etc.

Attracted by abilities in manipulating light on the wavelength or subwavelength scale, metasurfaces, especially metalens, has boosted a great range of applications, including optical trapping[10], full-color achromatic imaging[11-13], optical multi-parameters detection[14, 15] or multiphoton quantum source[16]. In this article, we propose and numerically demonstrate a design of niobium pentoxide ($Nb_2O_5$) integrated scalable metalens array (ISMA) for optical addressing in trapped-ion QC with the working wavelength of UV 350nm. The $Nb_2O_5$ ISMA can focus collimated the individual addressing beams array into chain-arranged diffraction-limited focus spots array, with spot interval of 5μm, crosstalk below 0.82%, focusing efficiency as high as 36.90% at working distance (WD) of about 30μm. For various applications, the x-polarized, polarization-insensitive and right-handed circularly polarized ISMA are designed and verified with excellent polarization maintaining properties. Our scheme might help trapped-ion QC scale to thousands of qubits.

2. Method and Design

The principle of the $Nb_2O_5$ ISMA is Illustrated in Fig.1. As the collimated individual addressing beams array normally incident on the ISMA, the ISMA can output chain-arranged focus spots array with uniform spot interval. Each focus spot will be accurately aligned to one exact trapped ion when applied to optical addressing. Fig.1(b) shows that the ISMA consists of periodical metalens moleculars with a period of d×5, where the d denotes the spot interval, n denotes the number of metalenses in one metalens molecular. As is depicted in Fig.1 (a), a metalens molecular is composed of five metalenses spatially centrosymmetricly arranged in the "Z" shape, and each metalens corresponds to one exact focus spot. Here we take the center of metalens 3 as the coordinate origin, the arrangement direction of the focus spots array as the x axis, and the direction of light propagation as the z axis. Furthermore, the two centrosymmetric metalenses 1 and 5 or 2 and 4 have the opposite values of off-axis lengths along the x and the y axis, respectively.

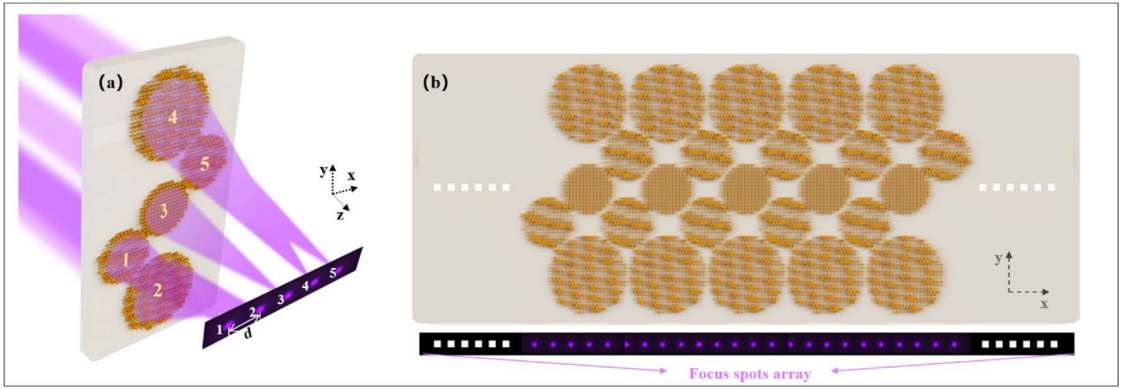

Fig.1. Schematic of the principle. (a) Diagram of a metalens molecular. (b) Diagram of the $Nb_2O_5$ ISMA. The inset shows the scalable focus spots array.

To enhance the design freedom of the $Nb_2O_5$ ISMA, the metalens adopts the hyperbolic off-axis phase distribution:

$$\varphi^{(i)}(x,y) = \frac{2\pi}{\lambda}\left[f^{(i)} - \sqrt{\left(x - d_x^{(i)}\right)^2 + \left(y - d_y^{(i)}\right)^2 + d_z^{(i)2}}\right] + \varphi_c \quad (1)$$

$$f^{(i)} = \sqrt{\left(d_x^{(i)}\right)^2 + \left(d_y^{(i)}\right)^2 + \left(d_z^{(i)}\right)^2} \quad (2)$$

Where $\varphi^{(i)}$ is the phase shift of the $i_{th}$ metalens, (x,y,0) is a certain point on the metalens plane, $f^{(i)}$ is the focal length of the $i_{th}$ metalens, $(d_x^{(i)}, d_y^{(i)}, d_z^{(i)})$ is the location of the $i_{th}$ focus, $d_z^{(i)}$ represents working distance, $\lambda$ is the working wavelength, and $\varphi_c$ is a constant reference phase. By optimizing the three parameters $d_x^{(i)}, d_y^{(i)}, d_z^{(i)}$, the position and optical performance of the focus spot can be flexibly manipulated. Note that when $d_z^{(i)}=0$, it is a traditional design of on-axis metalens.

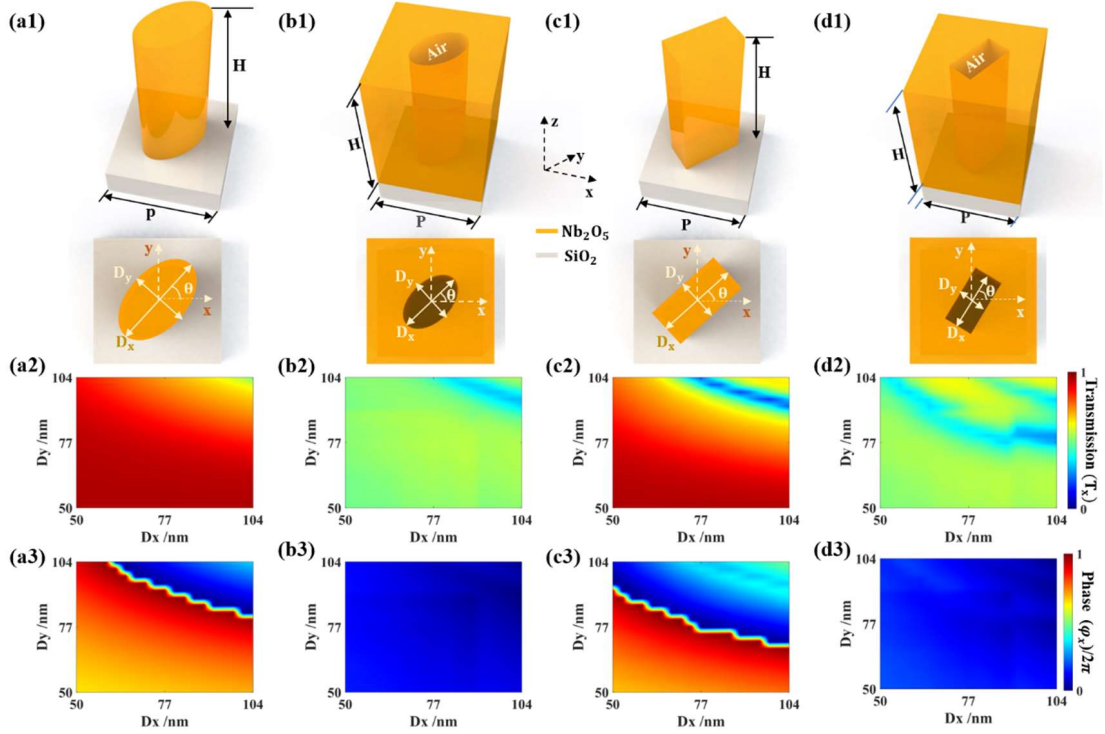

Fig.2. The simulation results of the four geometric types of nanopillars under the incidence of the horizontally linear polarized (x-polarized) light. (a1) (b1) (c1) (d1), Schematic of the ellipse, the inverse ellipse, the rectangle and the inverse rectangle pillar, respectively. (a2) (b2) (c2) (d2) The transmittances $T_x$ corresponding to the four types of the nanopillars. (a3) (b3) (c3)(d3) The phase $\varphi_x$ related to the four types of the nanopillars. *P, H*: the period and height of the nanopillars. $D_x$, $D_y$: the length of long and short axis (length and width) of the dielectric or air nanopillars. $\theta$: the rotation angle of the long axis (length of rectangle) relative to the positive direction of x axis

Here, as shown in Fig.2 (a1)-(d1), the nanopillars are composed of niobium pentoxide $Nb_2O_5$ pillar based on a square quartz substrate. The refractive index of $Nb_2O_5$ is 2.76+0.02*i*. To fully cover the phase shift range 0-2π, four geometric types (namely the ellipse, the inverse ellipse, the rectangle and the inverse rectangle pillar) are used to create a nanopillars library. The FDTD method is used to calculate the phase and transmittance of the nanopillars. In detail, the sweep range of the long and short axis is 50nm-104nm with the step length of 3nm, the period *P*=180nm, height *H*=350nm. Meanwhile, Boundary conditions are set similarly as in the reference[14, 17]. Fig. 2 (a2)-(d3) are the results of the transmittances $T_x$ and phase $\varphi_x$ under the incidence of x-polarized plane wave, it is very clear that single geometric type cannot cover the whole 0-2π, but the combinations of four geometric types can. Similarly, the transmittance $T_{rcp}$ and the phase $\varphi_{rcp}$ of the right-handed circularly polarized light can be obtained under the incidence of the left-handed circularly polarized light.

For the applications with different manipulations of polarization, the x-polarized sensitive (X), polarization-insensitive (PI), and the right-handed circularly polarized sensitive (RCP) ISMA are designed, respectively. Briefly, the X metalenses and the PI metalenses both follow the propagation phase principle, while the RCP metalenses are followed by the Pancharatnam-Berry (PB) phase principle[14, 18, 19]. For X metalenses, we select anisotropic nanopillars from the nanopillars library to constitute them with the smallest complex amplitude error compared with the ideal complex amplitude. For PI metalenses, isotropic nanopillars are selected in the similar way. However, for RCP metalenses, we deliberately select a rectangular pillar with $D_x$=104nm and $D_y$=59nm as the basic unit with the highest polarization conversion efficiency of 62.8%, as defined in the references[20].

For one spatially centrosymmetric metalens molecular of certain ISMA, only the individual metalens 2,3 and 5 are necessary to be designed and characterized. The design parameters of metalens 1 and 4 can be extracted from their centrosymmetric counterparts metalens 5 and 2. The detailed design parameters of metalens 2, 3 and 5 in the PI, X, RCP metalens molecular are listed in Table 1.

Table 1 Design parameters of the PI, X, RCP metalens molecular

| | Len number | Radius /μm | Position(x) /μm | Position(y) /μm | Off-axis length(x)/ μm | Off-axis length(y)/ μm | WD/ μm |
|---|---|---|---|---|---|---|---|
| **Polarization insensitive** | 2 | 12.5 | 0 | -27 | 5.1 | -27.4 | 30.9 |
| | 3 | 8 | 0 | 0 | 0 | 0 | 32.1 |
| | 5 | 8 | 12.5 | 10.5 | 2.6 | 10.9 | 31.7 |
| ↔ | 2 | 12.5 | 0 | -27 | 5.1 | -27.4 | 30.9 |
| | 3 | 8 | 0 | 0 | 0 | 0 | 31.4 |
| | 5 | 8 | 12.5 | 10.5 | 2.6 | 10.9 | 31.7 |
| ⟳ | 2 | 12.5 | 0 | -27 | 5.1 | -27.5 | 30.7 |
| | 3 | 8 | 0 | 0 | 0 | 0 | 31.4 |
| | 5 | 8 | 12.5 | 10.5 | 2.65 | 11.5 | 31.5 |

## 3. Characterization

To improve the simulation efficiency, the plane wave expansion is performed on the near-field distribution from the calculations by FDTD method, then the chirped Z transform is exploited to speed up the far-field calculation.

There are some metrics used in this paper. For clarity, Table 2 summarizes these definitions.

Table 2. Definitions of metalens's metrics

| Focusing efficiency | Crosstalk | Polarization extinction ratio |
|---|---|---|
| $I_{foc}/I_{in} \times 100\%$ | $I_N/I_{foc} \times 100\%$ | $I_{co\_max}/I_{cross\_max}$ |

**Spot radius:** the full-width at half-maximum of the x-slice of focal plane

$I_{foc}$: the intensity within the focus spot [diameter of 1.5μm] of the target metalens when the target metalens works and the others stop.

$I_{in}$: the incident intensity into the target metalens. when the target metalens works and the others stop.

$I_N$: the intensity within the focus spot [diameter of 1.5μm] of the target metalens as the target metalens stops while the others work.

$I_{co\_max}$: the maximum intensity of the target focus spot under the incidence of the co-polarization source when only the target metalens works.

$I_{cross\_max}$: the maximum intensity of the target focus spot under the incidence of the cross-polarization source when only the target metalens works.

Fig. 3 shows the simulation results of the PI, X, RCP metalens molecular. Fig.3 (a), (d), (g) are the normalized intensity distribution of xz cross-section for the PI, X, RCP metalens molecular, respectively. Fig.3 (b), (e), (h) are the normalized intensity distributions of the focal plane. Fig.3 (c), (f), (i) are the normalized intensity distribution of x-slice of focal plane. It is clear to see 5 chain-arranged focus spots array with consistent WD of around 30.00μm as well as uniform spots interval. There exist 4 elliptical focus spots (focus spot 1,2,4,5) in the focal plane of every metalens molecular own to the asymmetric phase distribution of corresponding metalenses. As for the overall performances, the X metalens molecular is superior to the PI metalens molecular for the smaller mean complex amplitude error. The reason is that there are more optional nanopillars for X metalens molecular, thus the mean complex amplitude errors of the X metalenses are smaller than those of the PI metalenses. Furthermore, the RCP metalens molecular surpasses the PI and X metalens molecular, including the highest brightness, smallest focus spots and highest focusing efficiency for each metalens. For example, the focusing efficiency of metalens 3 in the RCP, X, PI metalens molecular are 36.90%, 33.30%,31.00%, the crosstalk are 0.07%, 0.15%, 0.08%, respectively. This phenomenon is mainly due to the more uniform amplitude distribution and high polarization conversion efficiency of the rectangle nanopillars for RCP metalenses.

When it comes to the performance of one certain metalens molecular, we mainly focus on the RCP metalens molecular. As illustrated in Fig.3 (g)-(i), the focus spot 3 is brighter than focus spot 1 and 5 own to higher focusing efficiency (The focusing efficiency of metalenses 3 is 36.90%, 1 or 5 is 34.40%). It seems to be confusing that the metalens 3 is dimmer than the metalens 2, but the efficiency is higher. This is mainly a result that the area of metalens 3 is smaller than that of metalesns 2, therefor the incident intensity of the metalens 3 is less for the same irradiance. As we recall the definition of focusing efficiency, the focusing efficiency of metalens 3 is naturally higher. We further calculate the crosstalk (below 0.82%), the spot radius (0.61μm-0.75μm), and the polarization extinction ratio (above 133.61). The high polarization extinction ratios verify that the designed RCP metalens molecular maintains good fidelity. The results of the X and PI metalens molecular are similar.

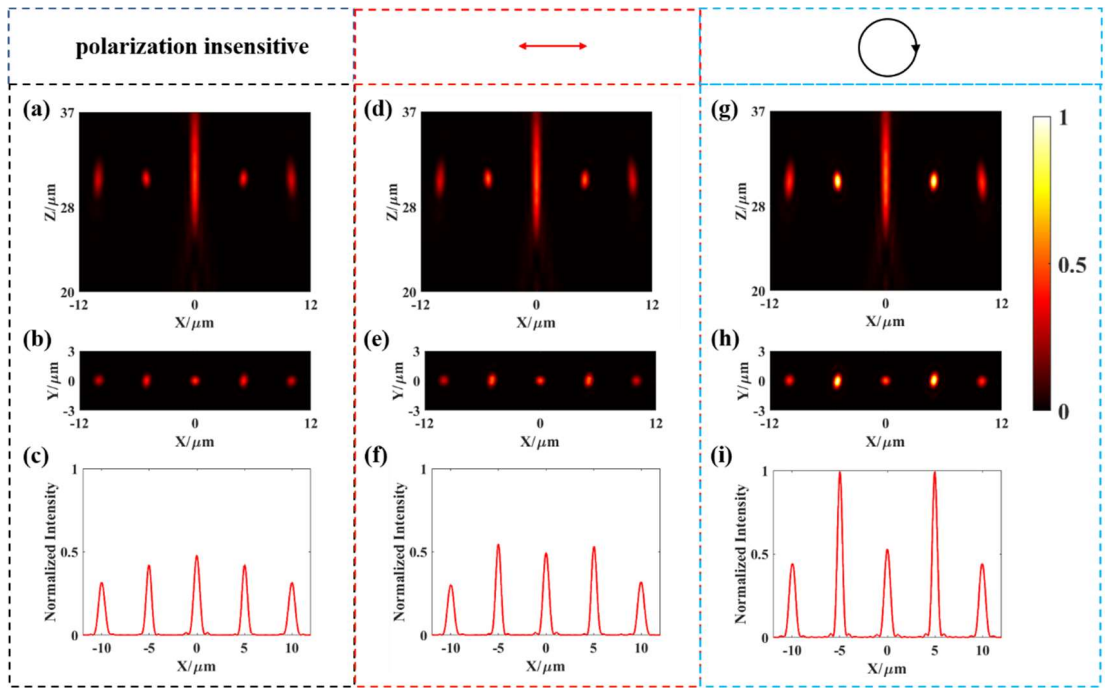

Fig. 3. Characterization of the PI, X and RCP metalens molecular. (a) (b) (c) Results of the PI metalens molecular. (d) (e) (f) Results of the X metalens molecular. (g) (h) (i) Results of the RCP metalens molecular. (a) (d) (g) The normalized intensity distribution of the xz cross-section. (b) (e) (h) The normalized intensity distribution of the focal plane. (c) (f) (i) The normalized intensity distribution of x-slice of the focal plane. All the intensity distributions are normalized by the maximum intensity of the RCP metalens molecular.

For simplicity, we just list the results of metalens 2, 3 and 5 of the PI, X and RCP metalens molecular in the Table3.

Table 3 Characterization of the PI, X and RCP metalens molecular

|  | Len number | Spot radius /μm | WD/μm | Crosstalk | Polarization extinction ratio | Focusing efficiency |
|---|---|---|---|---|---|---|
| polarization insensitive | 2 | 0.65 | 30.71 | 0.13% | 0.97 | 10.60% |
|  | 3 | 0.72 | 30.70 | 0.08% | 1.01 | 31.00% |
|  | 5 | 0.78 | 30.72 | 0.11% | 0.91 | 22.30% |
| 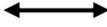 | 2 | 0.64 | 30.62 | 0.25% | 30.42 | 15.10% |
|  | 3 | 0.71 | 30.57 | 0.15% | 51.47 | 33.30% |
|  | 5 | 0.78 | 30.50 | 0.19% | 67.70 | 23.70% |
| 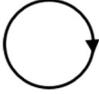 | 2 | 0.61 | 30.40 | 0.82% | 133.61 | 25.80% |
|  | 3 | 0.71 | 30.37 | 0.07% | 193.37 | 36.90% |
|  | 5 | 0.75 | 30.35 | 0.06% | 172.40 | 34.40% |

## 4. Discussion

We emphasize the comprehensive performance of the PI, x and RCP ISMA with their crosstalk below 0.82%, the uniform spot interval of 5μm, the large WD about 30.0μm, chain-arranged diffraction-limited focus spots array and high fidelity. The low focusing efficiency of the metalens array in this paper is mainly caused by the fact that the transmittance of niobium pentoxide ($Nb_2O_5$) thin film with a height of 350nm is only 53.50% calculated by FDTD method, and the metalenses with non-uniform amplitude distribution will perform extra amplitude modulation, which can deteriorate the efficiency of metalenses. The focusing efficiency can be improved by employing no-loss materials ($Si_3N_4, HfO_2$, etc.) as well as more geometric types of nanopillars with high transmittance. Furthermore, the design of ISMA can be customized according to specific goals. For example, if the required spot distance is 7μm, accordingly we can design an ISMA composed of metalenses with a diameter of 22μm. Due to the impressive dispersion manipulation capability of metasurfaces, we might design an multiwavelength or achromatic dielectric ISMA for optical addressing of trapped-ion QC, which can break the limit by the grating couplers.

## 5. Conclusion

In conclusion, we have designed the X, PI and RCP ISMA for optical addressing in trapped-ion QC, respectively, working in the violet 350nm range. The ISMA can convert two-dimensional individual addressing beams array into chain-arranged diffraction-limited focus spots array, with uniform spot distance of about 5μm, WD about 30μm, high fidelity, and crosstalk below 0.82%. Our study is vital to improve the scalability and integration of optical addressing in trapped-ion QC.